\documentclass[prb,twocolumn]{revtex4-1}

\usepackage{amssymb}
\usepackage{amsmath}
\usepackage[dvips]{graphicx}

\begin{document}

\title{Stability of magnetic nanoparticles inside ferromagnetic nanotubes}
\author{R. F. Neumann$^1$, M. Bahiana$^1$, J. Escrig$^{2,3}$, S. Allende$%
^{3,4}$, K. Nielsch$^5$, and D. Altbir$^{2,3}$}
\address{$^1$ Instituto de F\'isica, Universidade Federal do Rio de Janeiro, Caixa Postal
68528, Rio de Janeiro 21941-972, Brazil} 
\address{$^2$ Departamento de F\'isica, Universidad de Santiago de Chile (USACH),
Avenida Ecuador 3493, 917-0124 Santiago, Chile } 
\address{$^3$ Center for the Development of Nanoscience and Nanotechnology, CEDENNA,
917-0124 Santiago, Chile} 
\address{$^4$ Departamento de F\'isica, FCFM, Universidad de Chile, Casilla 487-3,
Santiago, Chile}
\address{$^5$ Institute of Applied Physics, University of Hamburg, Hamburg, Germany}

\begin{abstract}
During the last years great attention has been given to the encapsulation of
magnetic nanoparticles. In this work we investigated the stability of small
magnetic particles inside magnetic nanotubes. Multisegmented nanotubes were
tested in order to optimize the stability of the particle inside the
nanotubes. Our results evidenced that multisegmented nanotubes are more
efficient to entrap the particles at temperatures up to hundreds of kelvins.
\end{abstract}

\maketitle

Interest in magnetic nanostructures has increased dramatically over the past
decade, mainly due to the great progress in experimental techniques and
recent technological developments which allow access to interesting length
scales. In particular, understanding the behaviour of solids in confined
geometries at the nanoscale is an important area of basic research which has
a strong potential for applications \cite{LHH+09,Parkin09,FLH+06}. Within
this area, carbon nanotubes have stimulated intensive experimental and
theoretical research since their discovery in 1991 \cite{Iijima91}. One of
the topics that have been considered is the encapsulation of nanoparticles
for different purposes. In particular, paramagnetic nanoneedles have been
encapsulated \cite{KYG+05}, preventing them from agglomeration when a
magnetic field is not present and allowing control of the needles' movement.
The carbon encapsulation of ferromagnetic nanoparticles has also been
reported \cite{BLH07,RBS+10,KHK+08}, opening an interesting field of
research. In particular, encapsulation of Ni \cite{BZH+02}, Co \cite{BTX+02}%
, Fe \cite{FCW+07,MMP+10,SC08} and Ga \cite{LHX+09} inside multi-walled
carbon nanotubes has been reported and magnetically characterized using
vibrating sample magnetometers and superconducting quantum interference
devices. Also, the stability and magnetic properties of Fe encapsulated in
silicon nanotubes has been explored, finding a high local magnetic moment
per Fe atom, and a transition between from antiferromagnetic to a
ferromagnetic coupling is obtained by increasing the length of the nanotube 
\cite{WZM+07}.

Besides this intense study on the encapsulation of magnetic nanoparticles,
few attention has been given to the encapsulation of nanoparticles inside
magnetic nanotubes\cite{EBJ+08}. Magnetic nanotubes have been prepared since 2005 \cite{KNR+05} by K. Nielsch and his group. Also barcode-type nanostructures have been prepared. In particular, Lee \textit{et al.} \cite{LSN+05}  reported the synthesis of multisegmented metallic nanotubes with a bimetallic stacking configuration along the tube axis, showing a
different magnetic behavior as compared with continuous ones. 

Following these ideas, the focus of this work is the stability of a magnetic
nanoparticle inside a ferromagnetic nanotube. Using Monte Carlo simulations
we investigate different nanotubes, looking for the magnetic configuration
that increases the stability of an encapsulated magnetic nanoparticle. To
investigate the behaviour of a nanoparticle inside a magnetic nanotube we
start by calculating the magnetic field\cite{EAA+08} due to the nanotube
over the points of a three-dimensional grid $\left( \{\vec{r}_{i}\}\right) $%
. This step is accomplished by adding up the contributions to the overall
magnetic field coming from each atom that constitutes the nanotube (the
magnetic dipolar field). Then we simulate the random walk of a magnetic
nanoparticle over the field landscape generated by the nanotube. This is
done by means of a Monte Carlo simulation in which the nanoparticle starts
at the center of the nanotube and is allowed to move over the grid and to
rotate its magnetic moment under the influence of the tube's field.

The starting point is the definition of the crystalline structure of the
tube, i. e., we define the Bravais lattice, the lattice constant ($a_{%
\mbox
{\tiny 0}}$) and the magnetic moment per atom ($\mu _{\mbox {\tiny 0}}$).
The simulations shown in this paper consider BCC iron as the constituting
species, which gives us $a_{\mbox {\tiny 0}}=0.2866\ \mbox{nm}$ and $\mu _{%
\mbox{\tiny 0}}=2.22\ \mu _{\mbox{\tiny B}}$. An important point is the
definition of the three-dimensional grid over which the field will be
calculated and the energy of the particle will be tested. For our
calculations we have chosen a grid that is commensurate with the lattice
structure and that considers the available computational facilities.
According to these requirements, we have used as the grid step ($\Delta $),
the nearest multiple to $5\ \mbox{nm}$ of $a_{\mbox {\tiny 0}}$, hence, $%
\Delta =17\ a_{\mbox{\tiny 0}}=4.8722\ \mbox{nm}$. However, we have also
tried smaller and larger $\Delta $ ($\Delta =8\ a_{\mbox{\tiny 0}}$ and $52\
a_{\mbox{\tiny 0}}$, respectively) and no fundamental differences were
observed compared with the results obtained using $\Delta =17\ a_{%
\mbox{\tiny 0}}.$

The next step is to define the tube's overall dimensions. The tube's height,
which is also commensurate with $a_{\mbox {\tiny 0}}$, was chosen to be $L_{%
\mbox{\scriptsize tube}}=205\ \Delta =998.8\ \mbox{nm}$. The radial
dimensions were set to $R_{\mbox{\scriptsize e}}=50\ \mbox{nm}$ (external
radius) and $R_{\mbox{\scriptsize i}}=40\ \mbox{nm}$ (internal radius). We
investigate the behaviour of the particle using multilayer nanotubes
composed of magnetic and non-magnetic segments, as illustrated in Fig. 1.
The length of the non-magnetic segment, the spacer, has also been varied
(see Fig. 2) to obtain the desired field landscape, but similar results are
obtained if one uses a non optimal choice.

\begin{figure}[h]
\begin{center}
\includegraphics[width=8cm]{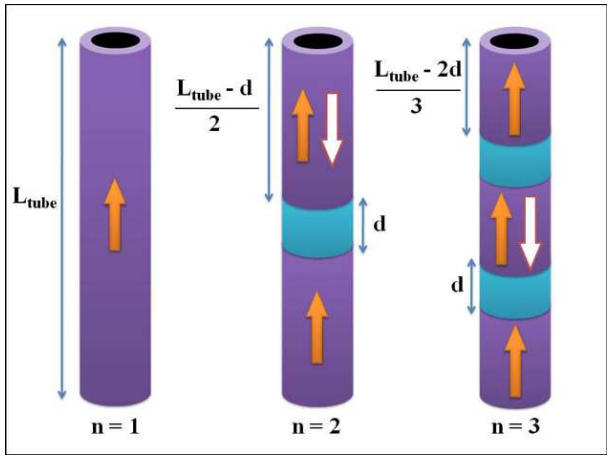}
\end{center}
\caption{(Color online) Illustration of the different tube geometries that
we have tested, depicting the magnetic and non-magnetic regions. The white
arrows represent what we call an antiparallel segments.}
\end{figure}

Even a nanotube with such dimensions would contain more than 240 million
atoms, rendering extremely expensive any attempt of direct calculations,
regardless of the computational power available. To circumvent this
technical issue, we first calculate the field generated over the grid points 
$\left( \{\vec{r}_{i}\}\right) $ by all the magnetic moments inside a ring ($%
\rho _{\mbox{\tiny 0}}$) of height $\Delta $ which is magnetized along its
axis and located at the centre of the grid. We used the usual expression
(Eq. \ref{B_ring}), for the field of a collection of magnetic dipoles and
stored our results for each point of the grid in a file. The latter was used
as a building block to construct the desired tube. 
\begin{multline}
\vec{B}_{0}\left( \vec{r}_{i}\right) =\sum_{a\ \in \ \rho _{\mbox{\tiny 0}}}%
\frac{3\vec{\mu}_{a}\cdot \left( \vec{r}_{i}-\vec{r}_{a}\right) }{|\vec{r}%
_{i}-\vec{r}_{a}|^{5}}\left( \vec{r}_{i}-\vec{r}_{a}\right)  \label{B_ring}
\\
-\frac{\vec{\mu}_{a}}{|\vec{r}_{i}-\vec{r}_{a}|^{3}}+\frac{8\pi }{3}\vec{\mu}%
_{a}\delta \left( \vec{r}_{i}-\vec{r}_{a}\right) ,
\end{multline}%
where the indices $a$ and $i$ run over all the atoms belonging to the ring $%
\rho _{\mbox{\tiny 0}}$ and all the sites belonging to the grid,
respectively.

Finally, by summing the contributions coming from different rings ($\rho _{n}
$) centered at different positions ($\vec{r}_{n}=n\Delta \hat{z}$) with
different weights ($w_{n}$), we are able to construct multi-segmented tubes 
\cite{LSL+09} with as many parallel or antiparallel aligned segments as we
want (just by weighting their contribution by $\pm 1$) and separated by
non-magnetic materials (by weighting their contribution by 0) as follows. 
\begin{equation}
\vec{B}_{\mbox{\scriptsize tube}}\left( \vec{r}_{i}\right)
=\sum_{n=-N}^{+N}w_{n}\vec{B}_{0}\left( \vec{r}_{i}-\vec{r}_{n}\right) ,
\label{B_tube}
\end{equation}%
where $N=\left( L_{\mbox{\tiny tube}}-\Delta \right) /2\Delta $. The result
of such calculation is depicted in Fig. 2, which shows the field along the
tube's axis obtained from Eq. 2 for one or two segments while keeping the
full length of the tube fixed at $998.8\mbox{ nm}$, as illustrated in Fig.
1. In this figure we can observe that for thinner spacers the antiparallel
alignment provides a higher magnetic field in the centre of the tube
compared to the parallel alignment. 

\begin{figure}[h]
\begin{center}
\includegraphics[width=8cm]{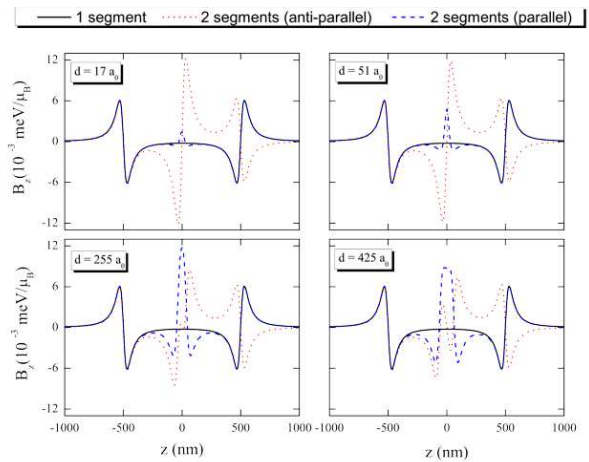}
\end{center}
\caption{(Color online) The magnetic field along the tube's axis obtained
from Eq. \ref{B_tube} for one or two segments while keeping the length fixed.}
\end{figure}

Above we have defined the set of grid points occupied by the tube itself and
the magnetic field $\vec B \left( \{ \vec r_i \} \right)$ that it creates.
So now we are in position to investigate the stability of a nanoparticle
inside the tube. We start by putting the particle, characterized by its
position $\left( \vec r\ \in \{\vec r_i\} \right)$ and magnetic moment
vector $\left( \vec \mu \right)$, in the center of the tube and let it move
and rotate under the Hamiltonian 
\begin{equation*}
\mathcal{H} = -\vec \mu \cdot \vec B \left(\vec r \right) \ .
\end{equation*}
The magnitude of the moment $\vec \mu$ is chosen to represent a Co
nanoparticle with diameter $\Delta$ whose magnetic moment initially points
in a random direction. At each Monte Carlo step, a movement and a new
orientation for the magnetic moment are proposed simultaneously. The
movement consists in walking one step on the grid along one of the six
possible directions $\left( +\hat{x},-\hat{x},+\hat{y},-\hat{y},+\hat{z},-%
\hat{z} \right)$. The new orientation is sampled randomly from a sphere with
radius $|\vec \mu|$.

The energy difference caused by the new position and orientation is
calculated and the Monte Carlo step is accepted (or not) by means of the
standard Metropolis algorithm at a given temperature $T$ \cite{BH02}. All
the simulations shown on this paper were done using a total of $10^{7}$
Monte Carlo steps. Then we repeat this procedure $2 \times 10^{3}$ times in
order to get the average walk of the particle on the magnetic field
landscape of the tube.

We have performed the above-mentioned procedure for different tube
geometries, magnetic configurations and temperatures. We have chosen the
geometry in which the non-magnetic spacer is $d = 51 a_{\mbox{\tiny 0}}$
thick because thinner spacers would allow long-range magnetic interactions
(such as RKKY) to play an undesired role by strengthening the coupling
between different segments. Our results are depicted in Fig. 3, which shows
a measure of the average z coordinate (along the tube's axis) of the
particles as a function of the number of Monte Carlo steps (mcs). The graphs
are shown in terms of $\xi = \sqrt{\langle z^{2}\rangle }/(L_{%
\mbox{\scriptsize tube}}/2)$. $\xi = 0$ denotes that particle is at the
center of the tube, $\xi \approx 0.34$ represents one of the interfaces
between the spacer and the magnetic section in the three-segment nanotube,
the tips of the tube are denoted by $\xi = 1$, and a particle leaving the
tube is denoted by $\xi > 1$.

Figure 3 shows that at $T=10\mbox{ K}$ all the compound tubes are able to
confine the particle while the simple tube is not. At $T=300\mbox{ K} $ none
of the tubes were capable of keeping the particle trapped, although the tube
with two antiparallel aligned segments was the one that mostly delayed the
leaving. However, it is at $T=100\mbox{ K}$ that interesting phenomena take
place: the parallel alignment of segments seems to be less effective in
trapping the particle, since only the field landscape created by
antiparallel aligned segments is capable of restraining the particle's
movement.

\begin{figure}[h]
\begin{center}
\includegraphics[width=8cm]{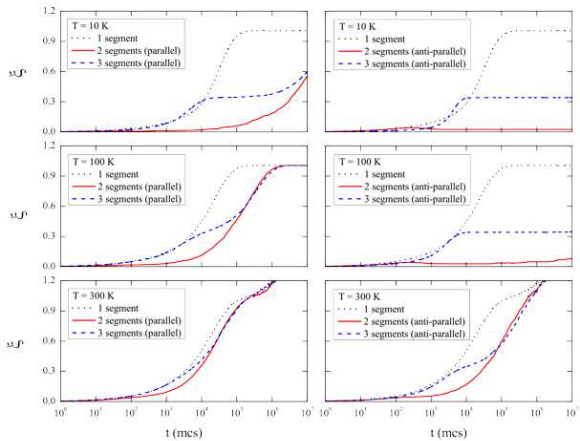}
\end{center}
\caption{(Color online) Evolution of $\xi $ as a function of the number of
Monte Carlo steps (mcs) for several tube geometries (one, two or three
segments), different magnetic configurations (parallel or antiparallel
alignment) and several temperatures ($T=10,\ 100,\ 300\mbox{ K}$). $\xi =1$
represents the tips of the tube.}
\end{figure}

We also investigate the effect of the size of the nanoparticle by increasing
its diameter from $\Delta $ to $2\Delta$. Our results for $T=300$ K are
presented in Fig. 4, showing that larger sizes contribute to stabilize the
particle inside the tube.

\begin{figure}[h]
\begin{center}
\includegraphics[width=8cm]{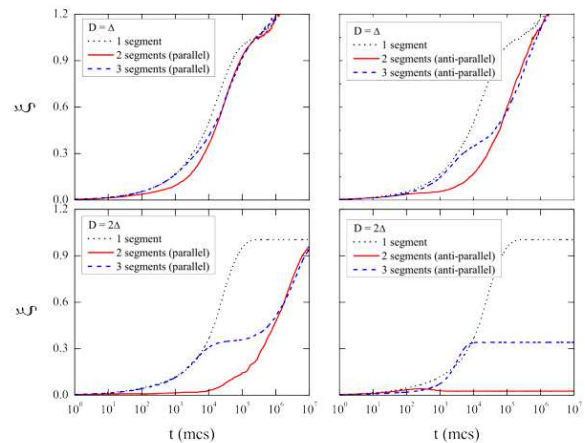}
\end{center}
\caption{(Color online) Evolution of $\xi $ as a function of the number of
Monte Carlo steps (mcs) for particles with diameter $\Delta $ (top) and $%
2\Delta $ (bottom).}
\end{figure}

From our results we can conclude that the use of multilayer geometries can
help to stabilize magnetic nanoparticles inside ferromagnetic nanotubes. In
spite there is still no experimental realization of these proposal, it might
be possible to synthetized magnetic nanoparticles inside nanotubes by
coating a multi-segmented nanotube with an oxide layer on the inner wall,
using Atomic Layer Deposition (ALD). Subsequently, using electrodeposition, the
nano-channel can be filled up to the middle of the pore channel with the
non-magnetic segment. At the end of the electrochemical deposition a short
segment of magnetic material can be deposited. Similar techniques have been
used to produce multilayered core/shell nanowires by Chong \textit{%
et al} \cite{CGM+10} and multi-segmented nanotubes by electrochemical deposition\cite{LSN+05}. Before starting the experiment, the non-magnetic metallic segment
is selective etched and the oxide layer should be also selectively removed. During the removal
of the oxide layer on the nanotube inner wall, and using a very strong
magnetic field in the perpendicular direction to the nanotube axis, it is
possible to trap and mechanically detach the magnetic particles from the template. By removing the perpendicular
magnetic field the time until the particle exit the tube can be measured.
Alternative it is possible to make the magnetic segments of a material with
a Curie temperature around 100 $^0$C and inject the magnetic particle with a
polymeric meld at a higher temperature into the channel. After cool down the
polymeric meld, it will solidify. The polymeric filling can be removed using
an organic solution and a strong magnetic field perpendicular to the
nanotube's diameter. In this way, these procedures can guide the design of
experiments which can test our results.

Looking at what happened at $T=100\mbox{ K}$, it seems possible to use the
alignment between the segments as a switch to open or close the particle's
exit. If one constructs a tube with two segments, one made of a soft and the
other of a hard magnetic material, one could in principle use an external
magnetic field to switch the direction of magnetization of the softer
segment, thus changing from antiparallel to parallel alignment and
facilitating the departure of the walking particle.

If one wants to increase the temperature where the free/trapped behaviour
occurs when using parallel/antiparallel alignment (in our case, $T=100%
\mbox{
K}$ for particles with diameter $\Delta$) one can simply use walking
particles with larger magnetic moments to increase the magnitude of the
Zeeman energy ($E_{\mbox{\tiny Zeeman}}=-\vec{\mu}\cdot \vec{B}$) with
respect to the thermal energy ($k_{\mbox{\tiny B}}T$).

In Chile we acknowledge support from FONDECYT under grants 1080300,
11070010, and 3090047, the Millennium Science Nucleus \textquotedblleft
Basic and Applied Magnetism\textquotedblright\ P06-022-F, and Financiamiento
Basal para Centros Cient\'{\i}ficos y Tecnol\'{o}gicos de Excelencia. In
Brazil we ackowledge PROSUL and CNPq. In Germany we acknowledge support from the German Research Council (DFG) in the framework of the Sonderforschungsbereich SFB668 Magnetismus vom Einzelatom zur Nanostruktur and by the Excellence Cluster \textquotedblleft Nanosprintronics\textquotedblright\ .

\end{document}